\documentclass[twoside]{article}
\usepackage{fleqn,psfig,espcrc2}
\usepackage[dvips]{graphics}
\usepackage[dvips]{color}
\usepackage{epsfig}
\usepackage{psfig}
\usepackage{lscape}
\usepackage{graphicx}

\newcommand{\bc}{\begin{center}}
\newcommand{\ec}{\end{center}}
\newcommand{\bd}{\begin{displaymath}}
\newcommand{\ed}{\end{displaymath}}
\newcommand{\bea}{\begin{eqnarray*}}
\newcommand{\eea}{\end{eqnarray*}}
\newcommand{\bi}{\begin{itemize}}
\newcommand{\ei}{\end{itemize}}

\newcommand{\beq}{\begin{equation}}
\newcommand{\eeq}{\end{equation}}
\newcommand{\beqn}{\begin{eqnarray}}
\newcommand{\eeqn}{\end{eqnarray}}

\chardef\galpha=11

\newcommand{\half}{\frac{1}{2}}

\newcommand{\J}[4]{{#1} {\bf #2} (#3) #4}

\newcommand{\NP}{Nucl.~Phys.}
\newcommand{\NPSup}{Nucl.~Phys.~B (Proc.~Suppl.)}
\newcommand{\PL}{Phys.~Lett.}
\newcommand{\PR}{Phys.~Rev.}

\definecolor{hinter1}{rgb}{0.90,0.91,0.96}
\definecolor{hinter2}{rgb}{0.96,0.93,0.90}
\definecolor{hinter3}{rgb}{0.90,0.96,0.91}
\definecolor{hinter4}{rgb}{0.5,0.5,0.5}
\definecolor{hinter5}{rgb}{1.00,0.92,0.95}

\definecolor{titletext}{rgb}{0.9,.1,.1}
\definecolor{titlebg}{rgb}{1.0,1.0,1.0}
\definecolor{zone1bg}{rgb}{0.96,0.93,0.90}
\definecolor{zone2bg}{rgb}{0.90,0.96,0.91}
\definecolor{zone3bg}{rgb}{0.8,0.94,1.0}
\definecolor{zone4bg}{rgb}{0.8,0.8,0.8}
\definecolor{background}{rgb}{0.4,0.4,0.4}

\definecolor{darkgreen}{rgb}{.1,.9,.1}
\definecolor{namecol}{rgb}{.1,.1,.9}
\definecolor{white}{rgb}{1.0,1.0,1.0}

\title{
\vspace*{-35pt}
Residual mass effects in improved domain wall fermions
\thanks{Presented by K.-I.Nagai at Lattice 2002.
KN is supported by JSPS Postdoctoral Fellowship for Research Abroad.
}
}

\author{
Pilar Hern\'andez\rlap,\address{CERN, Theory Division, CH-1211 Geneva 23, Switzerland}
Karl Jansen\address{NIC/DESY Zeuthen, Platanenallee 6, D-15738 Zeuthen, Germany} and
Kei-ichi Nagai$^{\rm a}$
}

\begin{document}
\setlength{\unitlength}{1cm}

\begin{abstract}
In order to improve simulations with domain wall fermions (DWFs),
it has been suggested to project out 
a number of low-lying eigenvalues
of the 4-dimensional Dirac operator 
that generates the transfer matrix of DWF. 
We investigate 
how this projection method affects 
chiral properties of quenched DWF.
In particular, 
we study the behaviour of the residual mass
as a function of the size of the extra dimension.
\end{abstract}

\maketitle

\section{Introduction}
\label{sec:intro}

Domain wall fermions (DWFs) preserve 
chiral symmetry \cite{kaplan,shamir,pert}
when the lattice size in the 5th direction, $N_s$, is taken to infinity.
A measure of chiral symmetry breaking is 
the residual mass $m_\mathrm{res}$. 
Even though the restoration of chiral symmetry is expected 
to be exponentially fast in $N_s$, 
in practice $m_\mathrm{res}$ decreases very slowly as
first shown by CP-PACS \cite{cppacs,eigenv}. 
This is due to the existence of very small
eigenvalues of the transfer matrix along the 5th direction. 
At large $N_s$ 
these eigenvalues determine the rate of the exponential decay. 
It is clear that any improvement of the chiral properties of DWF 
has to come from eliminating these low-lying eigenvalues.

One idea is 
to improve the gauge actions \cite{cppacs,eigenv,rbc,dbw2}. 
However, 
besides the potential difficulties with unitarity violations
and the sampling of topological charge sectors,
this method does not solve completely the problem. 
In particular 
with the Iwasaki gauge action, 
the convergence rate also becomes slow at large $N_s$ \cite{eigenv} 
(see Fig. \ref{fig:resmass}): very small eigenvalues of the transfer matrix, 
even though less frequently, also appear in this case. 
Using the DBW2 action seems to be much better in this respect \cite{dbw2}, 
but it is unclear whether these small eigenvalues could eventually
appear there, too, leading to similar problems.

Another method to eliminate the small eigenvalues is to
project them out of the transfer matrix \cite{project,boudp}.  
In this contribution,
we investigate the projection method based on Ref.\cite{project}, 
where the projection is performed on the transfer matrix itself.
(In Ref.\cite{boudp}, the projection is operated only on a boundary term.)
The aim of this paper is to investigate the effects of the projection 
method on the residual mass in the quenched configurations.

\section{Domain wall fermions}
\label{sec:dwf}

Details of our notation can be found in Ref.\cite{project}, 
and we give only a brief summary below.

The 5D domain wall operator is defined as
\beq
{\cal D} = \half \left\{ \gamma_5 \left( \partial^*_s + \partial_s \right) 
- a_5 \partial_s \partial_s \right\} + {\cal M} \quad ,
\label{eq:dwf}
\eeq
where the operator ${\cal M}$ is obtained from 
the standard 4D Wilson--Dirac operator 
with negative mass (DW mass).
Choosing open boundary conditions 
in the 5th dimension,
chiral modes with opposite chiralities 
are localized on the 4D boundary plane
at $s=1$ and $N_s$.
The action of ``quark fields'',
which are constructed from the boundary fermions,
is related to an effective 4D operator $D_{N_s}$, which satisfies
\beq
a D \equiv \lim_{N_s \rightarrow \infty} a D_{N_s} 
= 1 - \frac{A}{\sqrt{A^\dagger A }} \quad, 
\label{eq:gw} 
\eeq
where
\beq
A = \frac{- a_5 {\cal M}}{2 + a_5 {\cal M}} \quad.
\label{eq:matA}
\eeq
The $D$ of eq.(\ref{eq:gw}) satisfies the Ginsparg--Wilson relation
and thus an exact lattice chiral symmetry.
 
In realistic simulations,
for finite $N_s$,
chiral symmetry is explicitly broken to a certain extent 
that can be quantified  
by the values of the so-called residual mass,
which measures the anomalous term 
in the axial Ward--Takahashi identity \cite{pert}: 
$2 J_{5q}(x) \equiv \nabla_\mu A_\mu(x) - 2 m_f P(x)$, where 
$A_\mu$ is the axial current and $P$ is the pseudoscalar density.  
The size of this extra chiral symmetry-breaking term can be described by 
the residual mass
\beq 
m_\mathrm{res} = \lim_{t \rightarrow \infty}
\frac{\sum_{\bf x} \langle J_{5q}(t,{\bf x}) P(t,{\bf x}) \rangle }{\sum_{\bf x} \langle P(t,{\bf x}) P(t,{\bf x}) \rangle} \quad.
\eeq

\subsection{Improvement of domain wall fermion}
\label{subsec:imp}

We employ a method to                          
project out the small eigenvalue of $A^\dagger A$ \cite{project}.
The improved operator $\widehat{\cal{M}}$, instead of $\cal M$,
satisfies the following relations 
in order that eqs.(\ref{eq:gw}) and (\ref{eq:matA}) hold;
\beq
\widehat{\cal M}^\dagger = \gamma_5 \widehat{\cal M} \gamma_5 \quad, \quad
\det(2 + a_5 \widehat{\cal M}) \neq 0 \quad.
\label{eq:condition}
\eeq
This is used to construct $\widehat{\cal M}$
in such a way that the very low eigenvalues of $A^\dagger A$ disappear, 
while keeping $D$ invariant.
Following Ref.\cite{project},
the new operator ${\widehat {\cal M}}$ satisfying eq.(\ref{eq:condition})
is given by the following expression;
\beq
\widehat A = \frac{- a_5 \widehat{\cal M}}{ 2+ a_5 {\widehat{\cal M}}}
= A + \sum_{k=1}^r (\widehat \alpha_k - \alpha_k ) 
\gamma_5 v_k \otimes v_k^\dagger \,\, ,
\eeq
where $v_k$ is the eigenvector of $\gamma_5 A$,
\beq
\gamma_5 A v_k = \alpha_k v_k \,,\, \gamma_5 \widehat{A} v_k = \widehat{\alpha}_k v_k 
\,,\,
(v_k, v_l) = \delta_{k l} \,\,,
\eeq
and $r$ is the number of eigenvalues projected out.
Therefore an improved DWF operator 
can be obtained after substituting
${\cal M}$ in eq.(\ref{eq:dwf})
with $\widehat {\cal M}$ given by
\beq 
a_5 \widehat {\cal M} = a_5 {\cal M} - \sum_{k,l=1}^r
X_{kl} w_k \otimes w_l^\dagger \gamma_5 \quad,
\eeq 
where 
$(X^{-1})_{kl} = 2 \delta_{kl} (\widehat \alpha_k - \alpha_k)^{-1}
+ (v_k,w_l) $
and 
$w_k = (2 + a_5 {\cal M}) \gamma_5 v_k$.

\section{Eigenvalues of matrix $A^\dagger A$}
\label{sec:ev}

The eigenvalues of $A^\dagger A$ can be calculated by 
a straightforward generalization \cite{project} 
of the Ritz functional method in Ref.\cite{ritz}.
We have calculated the convergence rate 
from eigenvalues of $A^\dagger A$ \cite{project,impr}
and the distributions of low-lying eigenvalues 
for various gauge actions \cite{impr}.

\section{Simulation of improved DWF}
\label{sec:sim}

The values $\widehat{\alpha}_k$ can be chosen freely as long as 
$\widehat{\alpha}_k > \alpha_k$.
This can be used to reduce 
the value of the residual mass.
Here we use  
${\widehat \alpha}_k = 2 {\rm sgn}(\alpha_k) |\alpha_l|$
with $(\max\{k\} ; l)=(3 ; 10)$ and $(10 ; 10)$
in the plaquette gauge action.
The notation in the figures is 
$\lambda_\mathrm{proj}=2 \lambda_l = 2 |\alpha_l|$,
then ${\widehat \alpha}_k = {\rm sgn}(\alpha_k) \lambda_\mathrm{proj}$. 

For the Wilson gauge action we have chosen
the gauge coupling to be $\beta=6.0$ 
($a^{-1} \sim 2$ GeV) 
and the DW mass to be $m_0=1.8$.
The lattice size is $12^3 \times 24 \times N_s$.
This size is a little smaller than the ones used in simulations of 
the CP-PACS and RBC collaborations ($16^3 \times 32 \times N_s$);
the value of the lattice spacing is, however, the same.
The bare quark mass is always fixed to $m_f=0.02$.

\begin{figure}[h!tb] 
\centering
\centerline{\resizebox{8cm}{!}
{\rotatebox{-90}{\includegraphics{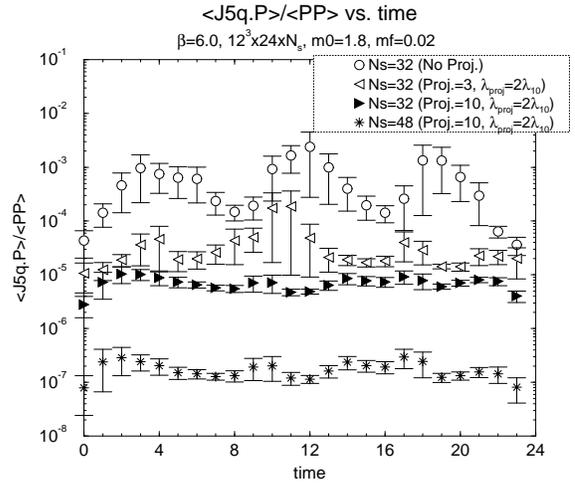}}}}
\caption{$\frac{\langle J_{5q}P \rangle}{ \langle PP \rangle}$
without and with projection for $N_s=32$ and with projection for $N_s=48$.
The fitting range for $m_\mathrm{res}$ is $4 \leq t \leq 20$.
}
\label{fig:projns32}
\end{figure}

In Fig.\ref{fig:projns32},
the quantity $\frac{\langle J_{5q}P \rangle}{ \langle PP \rangle}$
without and with projection for $N_s=32$ and 48 is plotted.
The number of projected modes is 0, 3 and 10.
Note the large fluctuations of the residual mass 
when no projection is performed.
As the number of eigenvalues projected out is increased,
the determination of the residual mass is much more stable
and cleaner.

\begin{figure}[h!tb] 
\centering
\centerline{\resizebox{8.5cm}{!}
{\rotatebox{-90}{\includegraphics{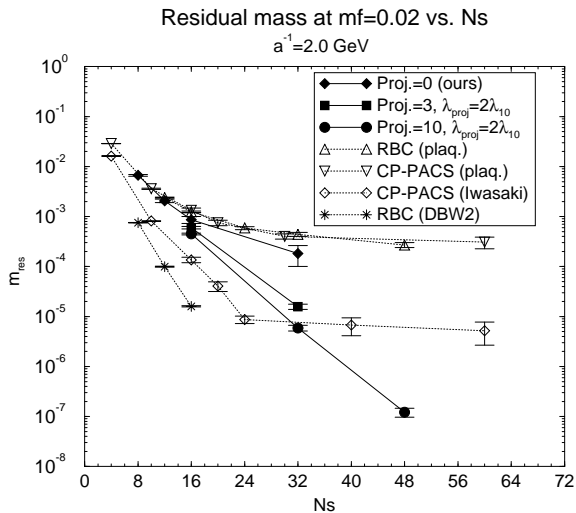}}}}
\caption{Residual mass as a function of $N_s$
with and without using the projection method.
Filled symbols represent our data.
We also add data from simulations of 
the CP-PACS and RBC collaborations for comparison.
The lines are just to guide the eyes.
}
\label{fig:resmass}
\end{figure}

Figure \ref{fig:resmass} shows our main result: 
the comparison of the behaviour of the residual mass 
against $N_s$ for different improvement methods. 
The result at small $N_s$, for example $N_s=16$, 
reveals that the projection method does not
show any effect. 
In the large-$N_s$ region, 
the effect of the projection of the low-lying eigenvalues 
on the residual mass is, however, remarkable.
This is explained \cite{eigenv} 
by simple and qualitative arguments, as follows;
$m_\mathrm{res} \sim \sum_{\alpha} e^{-\alpha N_s} 
\sim \int d \alpha \rho(\alpha) e^{-\alpha N_s}$,
where $\alpha$ is the square root of eigenvalues 
of the operator $A^\dagger A$ and $\rho(\alpha)$ is the eigenvalue density, 
which grows with $\alpha$.  
In the contribution to the residual mass of each eigenvalue, 
there is a competition between the density function 
and the exponential suppression.
Therefore, at large $N_s$,
the low-lying eigenvalues dominate owing to the exponential suppression.
At small $N_s$, however,
not only low-lying but also bulk eigenvalues contribute
to $m_\mathrm{res}$, owing to $\rho(\alpha)$.

\section{Summary}
\label{sec:summary}

We have studied the residual mass effect 
of the projection method \cite{project} 
on the restoration of 
chiral symmetry for quenched DWF. 
At large $N_s$,
the low-lying eigenvalues of $A^\dagger A$ dominate
the behaviour of the residual mass. 
Therefore the decay rate of the residual mass with $N_s$ 
can be simply controlled
by the number of modes projected out.
The residual mass for smaller $N_s$, however, is controlled by 
the low-lying and bulk eigenvalues.
This is consistent with the arguments in the previous section \cite{eigenv}.
Our numerical study shows 
that the projection method is superior 
to the improvement of the gauge action at large $N_s$. 

We have also observed that 
the quantity $\langle J_{5q}P \rangle / \langle PP \rangle$
becomes much more stable 
after performing the projection,
which may affect also other correlation functions
to be stable and clean.

Since the projection method leads to a small numerical overhead,
we conclude that using Wilson gauge action
combined with the projection method
is competitive with using improved gauge actions.
To understand the projection method further, 
we are investigating the projection method 
with improved gauge actions \cite{impr}.


\end{document}